# Superconductivity at 7.3 K in the 133-type Cr-based RbCr$_3$As$_3$ single crystals

Tong Liu[1,2], Qing-Ge Mu[1,2], Bo-Jin Pan[1,2], Jia Yu[1,2], Bin-Bin Ruan[1,2], Kang Zhao[1,2], Gen-Fu Chen[1,2,3], and Zhi-An Ren[1,2,3,*]

[1] Institute of Physics and Beijing National Laboratory for Condensed Matter Physics, Chinese Academy of Sciences, Beijing 100190, China

[2] School of Physical Sciences, University of Chinese Academy of Sciences, Beijing 100049, China

[3] Collaborative Innovation Center of Quantum Matter, Beijing 100190, China

* Corresponding Author: renzhian@iphy.ac.cn

**Abstract:**

Here we report the preparation and superconductivity of the 133-type Cr-based quasi-one-dimensional (Q1D) RbCr$_3$As$_3$ single crystals. The samples were prepared by the deintercalation of Rb$^+$ ions from the 233-type Rb$_2$Cr$_3$As$_3$ crystals which were grown from a high-temperature solution growth method. The RbCr$_3$As$_3$ compound crystallizes in a centrosymmetric structure with the space group of $P6_3/m$ (No. 176) different with its non-centrosymmetric Rb$_2$Cr$_3$As$_3$ superconducting precursor, and the refined lattice parameters are $a$ = 9.373(3) Å and $c$ = 4.203(7) Å. Electrical resistivity and magnetic susceptibility characterizations reveal the occurrence of superconductivity with an interestingly higher onset $T_c$ of 7.3 K than other Cr-based superconductors, and a high upper critical field $H_{c2}(0)$ near 70 T in this 133-type RbCr$_3$As$_3$ crystals.

Recently a new family of Cr-based superconductors $A_2Cr_3As_3$ (A = K, Rb, Cs) was reported and attracted lots of interests for the study of unconventional superconductivity in more 3d-transition-metal compounds [1-6]. These $A_2Cr_3As_3$ compounds have a particular quasi one-dimensional (Q1D) hexagonal crystal lattice with non-centrosymmetric symmetry, which is formed with infinite Q1D $(Cr_3As_3)^{2-}$ linear chains separated by alkali-metal cations. By replacing the $K^+$ ions with larger $Rb^+$ or $Cs^+$ ions, the superconducting $T_c$ decreases dramatically from 6.1 K to 4.8 K and 2.2 K respectively, which indicates distinct positive chemical pressure effect on its superconductivity upon lattice changing similar as the iron-based superconductors [7-9], although characterizations under external physical pressure on $K_2Cr_3As_3$ give monotonically reduced $T_c$ with increasing pressure [10, 11]. More interestingly, many experimental and theoretical research results point to intriguing spin-triplet paring superconductivity in these $A_2Cr_3As_3$ superconductors [4, 12-17]. Nevertheless, these $A_2Cr_3As_3$ compounds are extremely reactive when exposed in air and easily oxidized during most experimental procedures, which hinders many further studies for its intrinsic physical characteristics. Lately, by deintercalating half of the $A^+$ ions using ethanol from the $A_2Cr_3As_3$ lattice, another type of Q1D compounds $ACr_3As_3$ (A = K, Rb, Cs) with similar crystal structure but centrosymmetric lattice was obtained, while physical studies on these 133-type polycrystalline samples reported spin-glass ground state at low temperatures [18, 19]. However, by preparing single crystalline samples of the $KCr_3As_3$ and employing a delicate annealing process to improve the crystal quality, we recently found superconductivity with an onset $T_c$ of 5 K in this air-stable 133-type $KCr_3As_3$ [20].

In this letter we report the preparation and characterization of the 133-type Cr-based $RbCr_3As_3$ single crystal, which has an interestingly enhanced onset superconducting $T_c$ at 7.3 K.

The single crystals of $RbCr_3As_3$ were prepared by the deintercalation of $Rb^+$ ions from $Rb_2Cr_3As_3$ precursors, similar with our previous report of $KCr_3As_3$ [20]. At first, high quality single crystals of $Rb_2Cr_3As_3$ were grown out of RbAs and CrAs mixture using a high-temperature solution growth method [1]. Secondly, the as-grown

Rb$_2$Cr$_3$As$_3$ single crystals were immersed in pure dehydrated ethanol and kept for one week for the fully deintercalation of Rb$^+$ ions at room temperature. The obtained samples were washed by ethanol thoroughly. To further improve the sample quality, the as prepared crystals were annealed in an evacuated quartz tube at 373 K for 10 h [20]. All the experimental procedures were performed in a glove box filled with high-purity Ar gas to avoid introducing impurities. The obtained RbCr$_3$As$_3$ crystals are quite stable in air at room temperature.

The crystal structure of the RbCr$_3$As$_3$ single crystals was characterized at room temperature with a Bruker single crystal X-ray diffractometer using Cu-K$_\alpha$ radiation. The electrical resistivity was measured in a Quantum Design physical property measurement system (PPMS) by the standard four-probe method, and the dc Magnetization was measured by a Vibrating Sample Magnetometer system (VSM) under zero-field-cooling (ZFC) and field-cooling (FC) modes respectively.

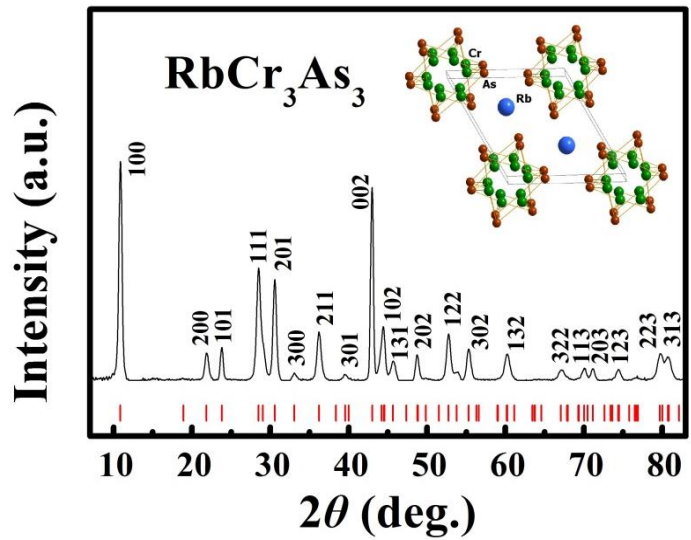

Fig. 1. The XRD patterns for the RbCr$_3$As$_3$ samples, and the vertical bars represent the position of calculated Bragg peaks. The insert depicts the crystal structure of RbCr$_3$As$_3$.

The XRD patterns collected from the RbCr$_3$As$_3$ samples at room temperature are presented in Fig. 1. All the diffraction peaks are well indexed with the hexagonal space group $P6_3/m$ (No. 176) as same as previously reported 133-type compounds [19], and no other impurity phase is detected in the diffraction peaks. The vertical bars represent the theoretical calculations for Bragg peak positions of the Rb-133 phase. The crystal structure was refined by the least-square fit method, which gave the lattice parameters a = 9.373(3) Å and c = 4.203(7) Å. Compared with the $Rb_2Cr_3As_3$ precursor, the crystal lattice of $RbCr_3As_3$ shows an obvious shrinkage along the a-axis while little change along c-axis by the ion deintercalation process. The schematic crystal structure of the Q1D hexagonal $RbCr_3As_3$ lattice is depicted as an inset in Fig. 1. We note that the energy-dispersive X-ray spectroscopy (EDS) measurements on the crystal surface of $RbCr_3As_3$ crystals give the atomic ratio of Rb:Cr:As close to 1:3:3, which also verifies the chemical composition of the obtained samples.

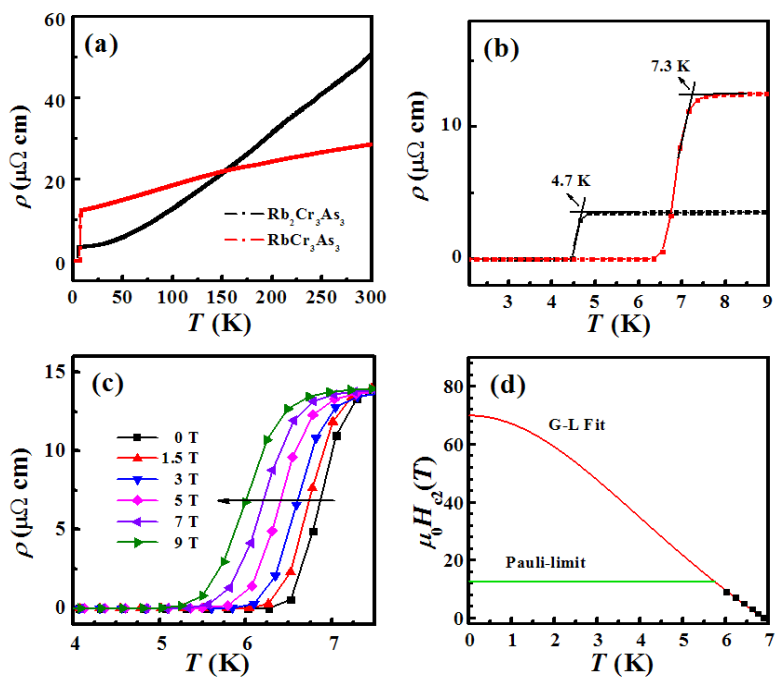

Fig. 2. (a) Temperature dependence of electrical resistivity for single crystals of $Rb_2Cr_3As_3$ and $RbCr_3As_3$. (b) Enlarged view for the resistive superconducting transitions. (c) The resistive transitions under different magnetic fields for $RbCr_3As_3$ from 0 T to 9 T. (d) Derived upper critical field and the Pauli paramagnetic limit for $RbCr_3As_3$.

The temperature dependence of electrical resistivity was characterized from 1.8 K to 300 K and the data are shown in Fig. 2. Both resistivity curves of the $Rb_2Cr_3As_3$ and $RbCr_3As_3$ show metallic behavior at the measuring temperature range, and superconducting transitions are observed in both samples at low temperatures, which is in contrast with previous report of spin-glass ground state for the polycrystalline Rb-133 sample [18]. The residual resistance ratio (RRR) of the RbCr$_3$As$_3$ (~ 2.3) is much smaller than that of Rb$_2$Cr$_3$As$_3$ (~ 15), indicating the still poor crystalline quality of the 133-type crystals mainly caused by the deintercalation process. Unexpectedly, for this 133-type RbCr$_3$As$_3$, the onset superconducting $T_c$ is enhanced up to 7.3 K, which is much higher than that of the 233-type Rb$_2$Cr$_3$As$_3$ (~ 4.7 K), and even higher than all other known Cr-based superconductors. The higher $T_c$ of Rb-133 than K-133 indicates a negative chemical pressure effect on $T_c$ in these 133-type superconductors which is opposed to that of the 233-type superconductors. The superconducting transition width is less than 1 K for both samples. To estimate the upper critical field, we performed resistivity measurements with sweeping temperature under constant magnetic fields from 0 T to 9 T (with the field perpendicular to the *c*-axis and electrical current along the *c*-axis), and the data are shown in Fig. 2c. With magnetic field increasing, the superconducting transitions shift to lower temperatures systematically. We define the $\mu_0H_{c2}$ as the field determined by 50% of the normal-state resistivity at $T_c$, and it was depicted as a function of temperature in Fig. 2d. Considering Ginzburg-Landau (GL) theory, $\mu_0H_{c2}(T) = \mu_0H_{c2}(0)(1 - t^2)/(1 + t^2)$, here $t = T/T_c$, the zero-temperature upper critical field $\mu_0H_{c2}(0)$ is estimated to be 70 T. This value is much higher than the Pauli paramagnetic limited upper critical field $\mu_0H_p = 1.84T_c = 12.6$ T [21], which gives an evidence for unconventional superconductivity in this RbCr$_3$As$_3$, and the quite large $\mu_0H_{c2}$ also indicates possible potentials for future high field applications.

The temperature dependence of magnetic susceptibility for the RbCr$_3$As$_3$ single crystal samples was characterized under a magnetic field of 20 Oe (perpendicular to the c-axis) and shown in Fig. 3. The results indicate clear diamagnetic superconducting transitions with an onset $T_c$ ~ 6.2 K. The shielding volume fraction at 2 K from the ZFC data is close to 100%, while the Meissner volume fraction from the FC data is quite small possibly due to the extremely strong flux pinning effect by remaining atomic defects inside the crystal lattice produced during the Rb$^+$ ions deintercalation process.

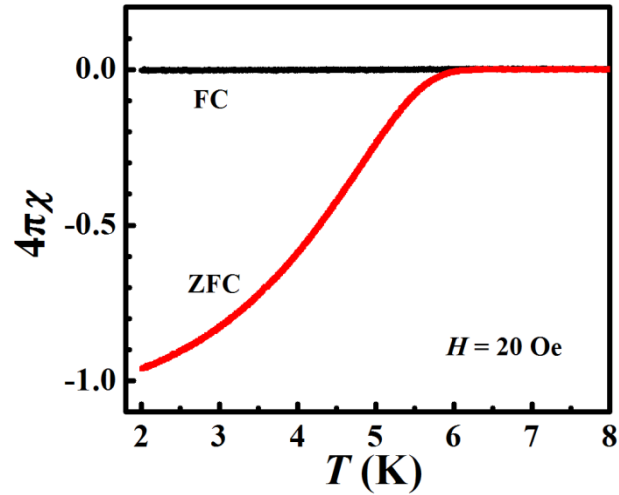

Fig. 3. Temperature dependence of magnetic susceptibility for the RbCr$_3$As$_3$ single crystals

In conclusion, we successfully prepared the 133-type RbCr$_3$As$_3$ single crystals, and the resistivity and magnetization measurements show clear superconducting transitions with a higher onset temperature of $T_c$ ~ 7.3 K than all other Cr-based superconductors, and a high upper critical field $H_{c2}(0)$ near 70 T. These 133-type superconductors also show a negative chemical pressure effect on $T_c$ in contrast to the 233-type superconductors.

**Acknowledgments**:

The authors are grateful for the financial supports from the National Natural Science Foundation of China (No. 11474339), the National Basic Research Program of China (973 Program, No. 2016YFA0300301) and the Youth Innovation Promotion Association of the Chinese Academy of Sciences.